\documentclass[iop,apj]{emulateapj}  %typeset as for main journal

\usepackage{epsfig}
\usepackage{amsmath}
\usepackage{amssymb}
\usepackage{natbib}
\usepackage{threeparttable} 

\usepackage{epstopdf}

\newcommand{\chan}{\textit{Chandra}}
\newcommand{\swift}{\textit{Swift}}
\newcommand{\xmm}{\textit{XMM-Newton}}

\newcommand{\Msun}{\mathrm{M}_{\sun}}
\newcommand{\Mearth}{\mathrm{M}_{\earth}}
\newcommand{\lum}{\mathrm{erg~s}^{-1}}
\newcommand{\flux}{\mathrm{erg~cm}^{-2}~\mathrm{s}^{-1}}
\newcommand{\cnts}{\mathrm{c~s}^{-1}}
\newcommand{\nh}{\mathrm{cm}^{-2}}

\newcommand{\sgra}{Sgr~A$^{*}$}

\slugcomment{Submitted to the Astrophysical Journal}
\shorttitle{Sgr~A$^{*}$ flares seen with Swift}
\shortauthors{Degenaar et al.}

\begin{document}

\title{The X-ray flaring properties of Sgr~A$^{*}$ during six years of monitoring with Swift}

\author{N. Degenaar$^{1,}$\altaffilmark{5}, J. M. Miller$^{1}$, J. Kennea$^{2}$, N. Gehrels$^{3}$, M. T. Reynolds$^{1}$, and R. Wijnands$^{4}$}
\affil{$^1$Department of Astronomy, University of Michigan, 500 Church Street, Ann Arbor, MI 48109, USA; degenaar@umich.edu\\
$^2$Department of Astronomy and Astrophysics, 525 Davey Lab, Pennsylvania State University, University Park, PA 16802, USA\\
$^3$Astrophysics Science Division, NASA Goddard Space Flight Center, Greenbelt, MD, USA\\
$^4$Astronomical Institute "Anton Pannekoek," University of Amsterdam, Postbus 94249, 1090 GE Amsterdam, The Netherlands}

\altaffiltext{5}{Hubble fellow}

%%%%%%%%%%%%%%%
% ABSTRACT
%%%%%%%%%%%%%%%

\begin{abstract}

\end{abstract}
\begin{abstract} 
Starting in 2006, \swift\ has been targeting a region of $\simeq 21' \times21'$ around Sagittarius A$^*$ (\sgra) with the onboard X-ray telescope. The short, quasi-daily observations offer an unique view of the long-term X-ray behavior of the supermassive black hole. We report on the data obtained between 2006 February and 2011 October, which encompasses 715 observations with a total accumulated exposure time of $\simeq$0.8~Ms. A total of six  X-ray flares were detected with \swift, which all had an average 2--10 keV luminosity of $L_{\mathrm{X}} \simeq (1-3) \times 10^{35}~\lum$ (assuming a distance of 8 kpc). This more than doubles the number of such bright X-ray flares observed from \sgra. One of the \swift-detected flares may have been softer than the other five, which would indicate that flares of similar intensity can have different spectral properties. The \swift\ campaign allows us to constrain the occurrence rate of bright ($L_{\mathrm{X}} \gtrsim 10^{35}~\lum$) X-ray flares to be $\simeq$0.1--0.2 per day, which is in line with previous estimates. This analysis of the occurrence rate and properties of the X-ray flares seen with \swift\ offers an important calibration point to asses whether the flaring behavior of \sgra\ changes as a result of its interaction with the gas cloud that is projected to make a close passage in 2013.
\end{abstract}


%%%%%%%%%%%%%%%
% INTRODUCTION
%%%%%%%%%%%%%%%

\section{Introduction}
Understanding accretion onto supermassive black holes and the associated feedback to their environment lies at the basis of understanding their formation, growth and evolution, the chemical enrichment of the interstellar medium, galaxy evolution, and the formation of large scale structures in the Universe. Sagittarius A$^*$ (\sgra) is a supermassive black hole that forms the dynamical center of the Milky Way Galaxy \citep[e.g.,][]{reid2004,ghez2008,gillessen2009}. It is the most nearby Galactic nucleus, located at $\simeq$8 kpc, and therefore allows for an unparalleled study of the fueling process of supermassive black holes.\footnote[6]{Throughout this work we adapt a distance of $D=8$~kpc towards \sgra\ \citep[see][and references therein]{genzel2010}.}

The bolometric luminosity of \sgra\ is $\simeq$8--9 orders of magnitude lower than expected for Eddington-limited accretion onto a supermassive black hole with a mass of $\simeq 4\times10^{6}~\Msun$. Its puzzling sub-luminous character has been explained in terms of a very low accretion rate, radiatively inefficient accretion flows, and outflows ejecting the in-falling matter \citep[for reviews, see][]{melia2001,genzel2010,morris2012}. Investigation of the fueling process of \sgra\ may help to understand a large class of dim galactic nuclei and to place these objects into context with brighter supermassive black holes \citep[e.g.,][]{ho1999,nagar2005}.

Although \sgra\ is currently quiescent, the detection of the Fermi-bubbles and time-variable fluorescent emission from nearby gas clouds suggests that it may have been more active and considerably brighter in the past \citep[e.g.,][]{sunyaev1993,koyama1996,muno2007,su2010,terrier2010,ponti2010,ponti2012}. Furthermore, it was recently discovered that an ionized gas cloud of $\simeq3~\Mearth$ is approaching the supermassive black hole and is projected to pass as close as $\simeq$2\,200~$R_s$ (where $R_s = 2GM/c^2$ is the Schwarzschild radius) around 2013 September \citep[][]{gillessen2012,gillessen2012_2}. The interaction with the gas cloud may enhance the activity of \sgra\ in the next few years--decades \citep[e.g.,][]{anninos2012,burkert2012,gillessen2012_2,moscibrodzka2012,schartmann2012}.

\begin{table*}
\begin{center}
\caption{\swift/XRT Monitoring Observations of the Galactic Center.\label{tab:obs}}
\begin{tabular*}{1.0\textwidth}{@{\extracolsep{\fill}}lcccccc}
\hline
\hline
Year & Start Date & End Date & Total \# Obs & Total Exposure Time & Cadence  & Average Exposure Time\\ 
 &  &  &  & (ks) & (obs~day$^{-1}$)  & (ks~obs$^{-1}$) \\ 
\hline
1 & 2006 Feb 24 & 2006 Nov 1 & 197 & 262 & 0.8 & 1.3 \\	
2 & 2007 Feb 17 & 2007 Nov 1 & 173 & 172 & 0.8 & 1.0\\ 
3 & 2008 Feb 19 & 2008 Oct 30 & 162 & 200 & 0.6 & 1.2 \\ 
4 & 2009 May 17 & 2009 Nov 1 & 39 & 40 & 0.2 & 1.0 \\ 
5 & 2010 Apr 7 & 2010 Oct 31 & 64 & 71 & 0.3 & 1.1 \\ 
6 & 2011 Feb 4 & 2011 Oct 25 & 80 & 76 & 0.3 & 1.0 \\ 
{\bf 1--6} & {\bf 2006 Feb 24} & {\bf 2011 Oct 25} & {\bf 715} & {\bf 821} & {\bf 0.5} & {\bf 1.1} \\ \hline
\end{tabular*}
\tablenotes{{\bf Notes.} This overview concerns XRT data obtained in the PC mode. In 2007 the monitoring observations were interrupted for 46 days between 2007 August 11 and September 26 \citep[][]{gehrels2007a,gehrels2007b}. 
}
\end{center}
\end{table*}

In the soft X-ray band, \sgra\ emits a steady X-ray luminosity of $L_{\mathrm{X}} \simeq 2 \times 10^{33}~\lum$ \citep[2--10 keV;][]{baganoff2001,baganoff2003}. However, occasionally the X-ray emission flares up by a factor of $\simeq$5--150 for tens of minutes to a few hours. To date, a few dozen of such X-ray flares have been detected by \chan\ and \xmm\ \citep[e.g.,][]{baganoff2001,baganoff2003,goldwurm2003,belanger2005,porquet2003,porquet08,trap2011,nowak2012}. Most observed flares have an intensity of $L_{\mathrm{X}}\lesssim10^{35}~\lum$ (2--10 keV), but on four occasions bright flares with $L_{\mathrm{X}}\simeq (1-5) \times 10^{35}~\lum$ have been detected \citep[][]{baganoff2001,porquet2003,porquet08,nowak2012}. The bright events can be modeled with a power law of photon index $\Gamma \simeq 2$, but the spectral slope of the weaker flares is less well constrained. 

The total duration and short-timescale variability observed during the X-ray flares suggests that the emission originates close to the black hole, within $\simeq$10$~R_s$. Therefore, the flares allow investigation of the inner accretion flow, and offer a new view of the accretion processes at work in the Galactic nucleus. Different mechanisms have been proposed as the origin of the flares, ranging from magnetic reconnection and electron acceleration models \citep[e.g.,][]{markoff2001,yuan2003,liu2004,liu2006}, to the infall of gas clumps or disruption of small bodies such as comets or asteroids \citep[e.g.,][]{cadez2006,tagger2006}. 

The proposed flare models usually invoke synchrotron or inverse Compton radiation processes. Apart from addressing the nature of the engine driving the flares, much effort has therefore gone into trying to understand the underlying emission mechanism, particularly by studying the flares simultaneously at different wavelengths \citep[][]{eckart2003,eckart2006,yusefzadeh2006,yusefzadeh2008,hornstein2007,aharonian2008,marrone2008,doddseden2009,trap2010,trap2011}. Constraining the repetition rate and spectral properties of the flares in the X-ray band is an important aspect of understanding the cause and emission mechanism of the flares \citep[e.g.,][]{porquet08,trap2010}. 

Our current knowledge about the X-ray flares from \sgra\ is based on observations with \chan\ and \xmm. However, the \swift\ satellite is also equipped with a sensitive X-ray telescope that provides sufficient spatial resolution (comparable to that of \xmm) to study the X-ray flares. In this work, we report on the results obtained during a six-year long monitoring campaign of the Galactic center carried out with \swift.

%%%%%%%%%%%%%%%%%%
% OBSERVATIONS + RESULTS
%%%%%%%%%%%%%%%%%%

\section{Description of the Swift Campaign}\label{sec:campaign}
The \swift\ observatory \citep[][]{gehrels2004} carries an X-Ray Telescope \citep[XRT;][]{burrows05}, that is sensitive in the 0.5--10 keV range, has a field of view (FOV) of $\simeq21'\times21'$, and a spatial resolution of $\simeq3''$--$5''$ (90\% confidence level). 
The XRT can be operated either in the photon counting (PC) mode, in which a 2-dimensional image is acquired, or the windowed timing (WT) mode, in which the CCD columns are collapsed into one dimension to reduce the frame time. During standard operations, the XRT automatically selects the best mode based on the brightness of sources in the field \citep[][]{hill2004}. 

In 2006 February, \swift\ embarked on a program to monitor the central $\simeq21' \times 21'$ of our Galaxy with the onboard XRT \citep[][]{kennea_monit,degenaar09_gc,degenaar2010_gc}. This campaign consists of short ($\simeq1$~ks), pointed X-ray observations that are carried out in the months February--October.\footnote[7]{The Galactic Center cannot be observed with \swift\ between November--January due to Sun-angle constraints.} In 2006--2008 the exposures were taken approximately once every day, whereas the cadence was lowered to one observation every $\simeq3$~days from 2009 onwards (Table~\ref{tab:obs}). 

We use the \swift\ monitoring campaign of the Galactic center to study the long-term X-ray behavior of \sgra. We restrict ourselves to the PC mode data, so that  spatial information is available. Between 2006 and 2011, a total of 715 observations were carried out, amounting to a total exposure time of $\simeq0.8$~Ms (Table~\ref{tab:obs}). Depending on the exact spectral model, the typical sensitivity in a 1-ks exposure is $L_{\mathrm{X}} \simeq 5 \times 10^{33}~\lum$ (2--10 keV) for a point source located in the Galactic center region. Single XRT observations usually consist of a number of short data segments with an exposure time of $\simeq$100--500~s, that are separated by one satellite orbit ($\simeq$1.5~h).

 \begin{figure*}
 \begin{center}
	\includegraphics[width=18.0cm]{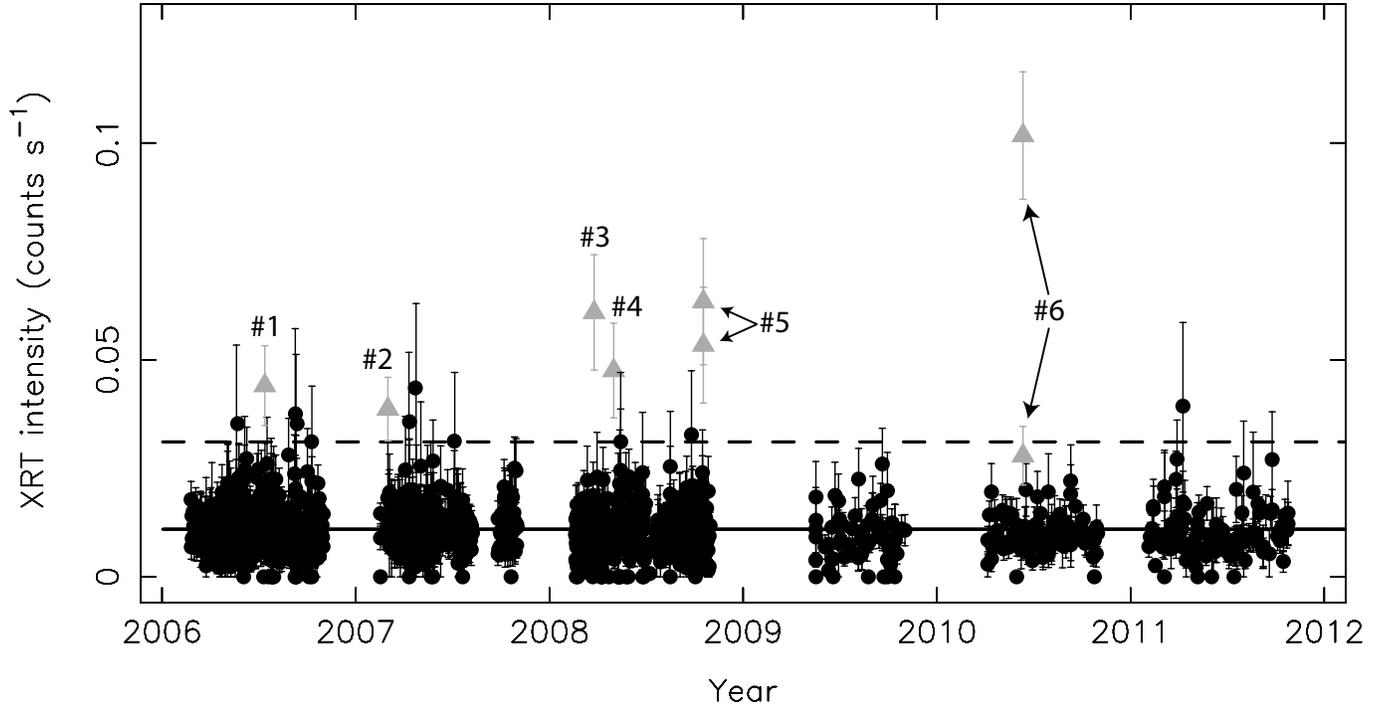}
    \end{center}
    \caption[]{Long-term XRT light curve of \sgra, binned per gti interval (0.3--10 keV). The solid horizontal line indicates the mean count rate observed in 2006--2011, whereas the dashed line indicates the 3$\sigma$ level. The six confirmed X-ray flares are numbered and indicated by light gray triangles (Table~\ref{tab:flares}). Flares \#5 and \#6 were both detected in two subsequent gti intervals. 
    }
 \label{fig:lc}
\end{figure*}

\section{Data Reduction and Analysis}\label{sec:ana}

\subsection{XRT Data Reduction}\label{subsec:reduction}
We processed all data of the Galactic center using the \swift\ tools ver. 3.8 within \textsc{heasoft} ver. 6.11, and using the calibration data released on 2011 August (ver. 3.8). As an initial step, all observations were re-processed using the tool \textsc{xrtpipeline}. For events of interest, X-ray spectra and light curves were extracted with \textsc{XSelect}. Ancillary response files (arfs) were created using the task \textsc{xrtmkarf}, taking into account exposure map corrections. The appropriate version of the response matrix files were obtained from the calibration data base (\textsc{caldb}). The X-ray spectral data was fitted within the \textsc{XSpec} package \citep[][]{xspec}. In all spectral fits we took into account interstellar absorption by using the TBABS model \citep[][]{wilms2000}.

Throughout this work we give the {\it absorbed} 2--10 keV fluxes ($F_{\mathrm{X}}$), because these are the least subject to systematic uncertainties in the spectral fitting procedure, and allow for a direct comparison with previously reported X-ray flares. Absorption-corrected luminosities ($L_{\mathrm{X}}$) were calculated by assuming a distance of $D=8$~kpc. Quoted errors for spectral parameters and fluxes refer to 90\% confidence intervals (in one interesting parameter). We note that all events detected from \sgra\ had count rates of $\lesssim0.2~\cnts$, and therefore pile-up is not an issue.

\subsection{Continuum X-ray emission around \sgra}\label{subsec:continuum}
To obtain X-ray light curves and spectra, we use a circular region with a radius of $10''$ centered on the radio position of \sgra\ \citep[][]{yusefzadeh1999}. \chan\ observations providing sub-arcsecond resolution suggest that a $10''$ extraction region does not only contain \sgra, but also contributions from (persistent) faint X-ray point sources and a strong diffuse emission component \citep[][]{baganoff2003,porquet08,muno2009}. 

We investigated the continuum X-ray emission detected at the position of \sgra\ to have a background reference for studying the X-ray flares (Section~\ref{subsec:specana}). To this aid we extracted accumulated spectra for each year of the campaign after removing possible flare events (Section~\ref{sec:results}). We fitted the spectral data with a simple absorbed power law model (TBABS*PEGPWRLW). We found that the spectral parameters and fluxes inferred for the different years were consistent within the errors. This indicates that the year-averaged sky emission detected by \swift\ is stable. 

A simultaneous fit to the year-averaged spectra with all parameters tied yields $N_{\mathrm{H}}=(9.1 \pm 0.5)\times10^{22}~\nh$, $\Gamma=2.5\pm0.1$, and an (absorbed) 2--10 keV flux of $F_{\mathrm{X}} = (2.1\pm 0.1)\times10^{-12}~\flux$. This is consistent with previous results obtained with \chan\ when using a similar extraction region \citep[][]{baganoff2003}. The quiescent emission of \sgra\ contributes only $\simeq10\%$ to the total flux observed within our XRT extraction region \citep[][]{baganoff2001,baganoff2003}.

\begin{table*}
\begin{center}
\caption{Basic Parameters of the X-Ray Flares Detected with \swift/XRT.\label{tab:flares}}
\begin{tabular*}{1.0\textwidth}{@{\extracolsep{\fill}}lccccccc}
\hline
\hline
\# & Obs ID & Date & Start Time & Duration & Counts & Rate & Significance  \\ 
 & &  & (hh:mm:ss.s) & (h) & &  ($10^{-2}~\cnts$) & ($\sigma$)  \\
\hline
1 & 35063119 & 2006 Jul 13 & 21:58:13.1 & $>$0.1 & $23$ & $4.4\pm0.9$ & 4.9  \\ 
2 & 30897001 & 2007 Mar 3 & 00:38:20.4 & $>$0.2 & $28$ & $3.9\pm0.7$ & 4.1  \\ 
3 & 35650045 & 2008 Mar 25 & 21:54:09.0 & $>$0.1 & $21$ & $6.1\pm1.3$ & 7.4  \\ 
4 & 35650068 & 2008 May 1 &18:57:14.1 & $0.1 < t < 6.2$ & $19$ & $4.8\pm1.1$ & 5.4  \\  
5 & 35650179 & 2008 Oct 17 &18:50:54.2 & $1.7 < t < 4.7$ & $35$ & $5.8\pm1.0$ & 7.5  \\ 
5a & \nodata & \nodata & \nodata & \nodata & $19$ & $6.5\pm 1.5$ & 7.8  \\  
5b & \nodata & \nodata & \nodata & \nodata & $16$ & $5.3\pm 1.3$ & 6.3  \\ 
6 & 90416021 & 2010 Jun 12 &10:22:59.6 & $>$1.7 & $65$ & $6.0\pm0.7$ & 10.5  \\ 
6a & \nodata & \nodata & \nodata & \nodata & $48$ & $10.2\pm 1.5$ & 13.5  \\  
6b & \nodata & \nodata & \nodata & \nodata & $17$ & $2.8\pm 0.7$ & 2.5 \\ 
\hline
\end{tabular*}
\tablenotes{{\bf Notes.} Flares \#5 and \#6 were both detected in two subsequent orbits. We give the number of counts, the count rate and the significance  both for the separate orbits (denoted by ``a'' and ``b''), as well as the values averaged over the two intervals. 
}
\end{center}
\end{table*}

\subsection{Searching for X-Ray Flares From \sgra}\label{subsec:searchflares}
To search the \swift\ data for X-ray flares, we created a long-term XRT light curve of \sgra. We extract the count rate for each single XRT good time interval (gti; generally these represent different satellite orbits, see Section~\ref{sec:campaign}). Occasionally, data segments of limited exposure were obtained, e.g., when an observation was interrupted by a trigger of \swift's Burst Alert Telescope, or due to an automated switch to the WT mode triggered by activity of one of the bright X-ray transients that are located in the FOV \citep[e.g.,][]{degenaar2010_gc,degenaar2012_xmmburst}. Since very short exposures do not generate reliable results, we chose to reject intervals with an exposure time of $<$100~s. After applying this selection, we were left with a total of 1349 gtis that were searched for occurrences of X-ray flares. The resulting long-term \swift/XRT light curve is shown in Figure~\ref{fig:lc}.

We searched the XRT light curve for occurrences of events that had an intensity of $>$3$\sigma$ above the mean and marked these as potential X-ray flares. We subsequently manually inspected the images and light curves of these observations to investigate whether these could indeed contain X-ray flares from \sgra\ (see Figures~\ref{fig:lcflare5} and~\ref{fig:imageflare} for examples). We analyzed the X-ray spectra of all events that were positively identified as X-ray flares and yielded sufficient source counts.

\subsection{Spectral Analysis of the X-Ray Flares}\label{subsec:specana}
The short XRT exposures typically provided a few tens of photons per event. Therefore we chose to fit the unbinned X-ray spectra without background subtraction using Cash statistics \citep[][]{cash1979}. We modeled the flares as a simple power law (PEGPWRLW), which was included as an additional component to the underlying continuum emission. All the parameters for the continuum were fixed to the values obtained from simultaneously fitting the year-averaged spectra (Section~\ref{subsec:continuum}), so that the index and normalization (i.e., the flux) of the flare spectrum were the only free fit parameters. The additional power law describing the flares is thus absorbed with $N_{\mathrm{H}}=1\times10^{23}~\nh$ fixed.

To allow for a model-independent test of the overall spectral shape of the X-ray flares, we also determined their hardness ratio ($HR$). For the present purpose we defined $HR$ as the ratio of counts in the 4--10 and 2--4 keV energy bands.

\begin{table*}
\begin{center}
\caption{Results from Spectral Analysis of the X-Ray Flares Detected with \swift/XRT.\label{tab:spec}}
\begin{tabular*}{1.0\textwidth}{@{\extracolsep{\fill}}lccccc}
\hline
\hline
\# & HR & $\Gamma$ & Cstat (dof) & $F_{\mathrm{X}}$ & $L_{\mathrm{X}}$  \\ 
 &  &  & & ($\flux$) & ($\lum$)  \\
\hline
1 & $2.1\pm1.0$ & $1.7\pm1.5$ & 18.89 (19) & $(1.0\pm0.6)\times10^{-11}$ & $(1.4\pm0.5)\times10^{35}$ \\ 
2 & $1.3\pm0.5$ & $1.9\pm1.5$ & 35.03 (26) & $(0.7\pm0.4)\times10^{-11}$ & $(0.9\pm0.3)\times10^{35}$ \\ 
3 & $2.5\pm1.2$ & $3.1\pm2.4$ & 21.62 (18) & $(1.5\pm0.9)\times10^{-11}$ & $(2.9\pm1.2)\times10^{35}$ \\ 
4 & $1.9\pm0.9$ & $1.8\pm1.8$ & 12.11 (16) & $(1.2\pm0.6)\times10^{-11}$ & $(1.5\pm0.9)\times10^{35}$ \\ 
5 & $1.9\pm0.7$ & $1.9\pm1.2$ & 29.23 (31) & $(1.3\pm0.5)\times10^{-11}$ & $(1.8\pm0.7)\times10^{35}$ \\ 
5a & $2.8\pm 1.5$ & $1.6\pm 1.6$ & 26.28 (16) & $(1.5\pm 1.1)\times10^{-11}$ & $(2.0\pm 0.9)\times10^{35}$ \\ 
5b & $1.3\pm 0.6$ & $2.7\pm 2.5$ & 14.06 (15) & $(1.0\pm 0.5)\times10^{-11}$ & $(1.8\pm 0.8)\times10^{35}$ \\ 
1--5 & $1.7\pm 0.2$ & $2.0\pm 0.6$ & 97.73 (105) & $(1.1\pm 0.3)\times10^{-11}$ & $(1.5\pm 0.3)\times10^{35}$ \\ 
6 & $0.7\pm0.2$ & $3.0\pm0.8$ & 50.52 (59) & $(0.9\pm0.2)\times10^{-11}$ & $(1.8\pm0.3)\times10^{35}$ \\ 
6a & $0.8\pm 0.2$ & $3.0\pm 0.9$ & 48.03 (43) & $(1.6\pm 0.4)\times10^{-11}$ & $(3.1\pm 0.5)\times10^{35}$ \\ 
6b & $0.5\pm 0.4$ & $3.2\pm 2.6$ & 11.23 (13) & $(3.4\pm 2.0)\times10^{-12}$ & $(0.7\pm 0.3)\times10^{35}$ \\ 
\hline
\end{tabular*}
\tablenotes{{\bf Notes.} The spectral data of the X-ray flares were fitted using an absorbed power law model with $N_{\mathrm{H}}=1\times10^{23}~\nh$ fixed, and applying Cash statistics. The hardness ratio ($HR$) is defined as the ratio of counts in the 4--10 and 2--4 keV bands. $F_{\mathrm{X}}$ gives the {\it observed} (i.e., not corrected for absorption) flux in the 2--10 keV band. $L_{\mathrm{X}}$ represents the average 2--10 keV luminosity corrected for absorption and assuming $D=8$~kpc. Flares \#5 and \#6 were both detected in two subsequent orbits. We give the spectral and color information both for the separate orbits (denoted by ``a'' and ``b''), as well as the values averaged over the two intervals. 
}
\end{center}
\end{table*}

\section{Results}\label{sec:results}
The mean XRT count rate of \sgra\ inferred from the 2006--2011 data set is $1.1\times10^{-2}~\cnts$, with a standard deviation of $6.7\times10^{-3}~\cnts$. The long-term light curve contained six events that had an intensity of $>$4$\sigma$ above the mean, and collected enough photons to allow for further study. Investigation of the images, light curves and spectra suggests that all six can be identified as X-ray flares from \sgra. These are numbered and indicated by light gray triangles in Figure~\ref{fig:lc}. 

\begin{figure}
 \begin{center}
		\includegraphics[width=8.8cm]{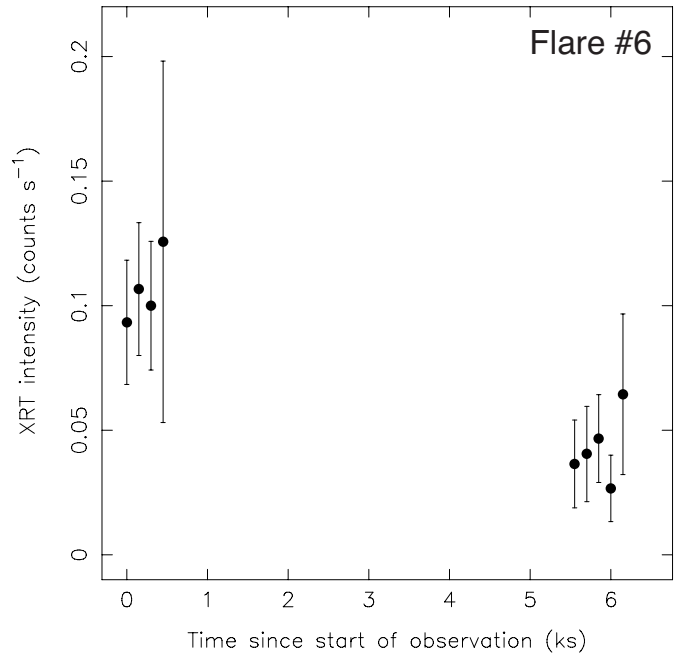}
    \end{center}
    \caption[]{
    \swift/XRT light curve of the flare that was observed on 2010 June 10 using a bin time of 120~s (0.3--10 keV). The flare was detected during both satellite orbits.
    }
 \label{fig:lcflare5}
\end{figure}

The basic properties of the six flares are summarized in Table~\ref{tab:flares}, while the results of the spectral analysis are presented in Table~\ref{tab:spec}. One of the flares (\#6) is briefly discussed in more detail in Section~\ref{subsec:flare6}. The observational setup (i.e., short exposures of $\lesssim1$~ks separated by $\gtrsim1$~h) of the \swift\ program does not constrain the full light curves of the flares (seeFigure~\ref{fig:lcflare5}). Therefore, we cannot determine their exact duration and peak intensity (Table~\ref{tab:flares}). 

In addition to the six confirmed X-ray flares, analysis of the long-term light curve revealed another ten individual events that had an intensity of a factor of $\simeq$3--4 above the mean. However, further investigation showed that these concerned short observations ($\simeq$100--200~s) that collected only a few counts ($\lesssim$10). Hence, these events were not statistically significant.

%%%%%%%%%%%%%%%
% BRIGHT FLARES
%%%%%%%%%%%%%%%

%%%%
% 6
%%%%

\subsection{Flare \#6: 2010 June 12}\label{subsec:flare6}
\begin{figure*}
 \begin{center}
	\includegraphics[width=8.8cm]{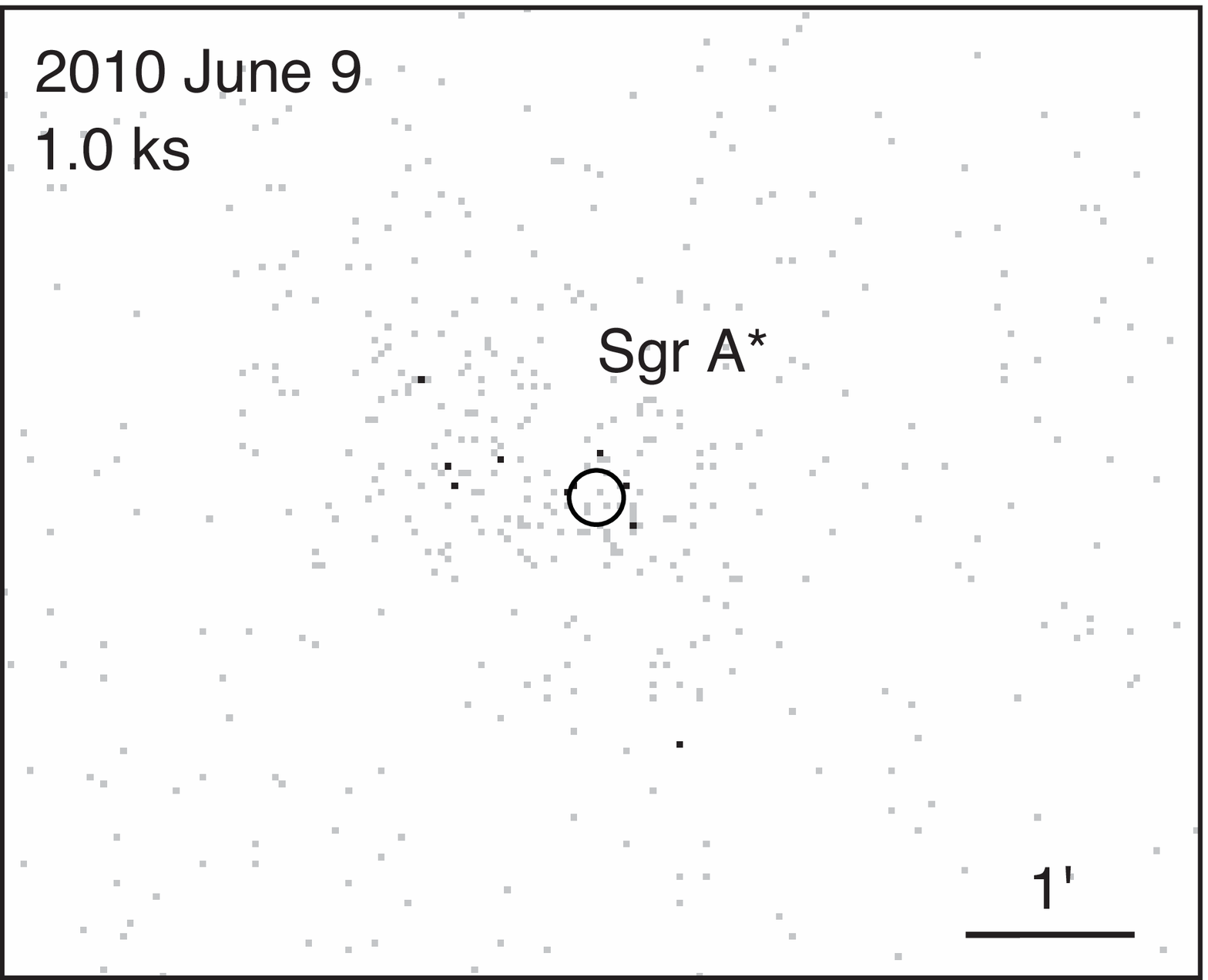}\hspace{0.2cm}
		\includegraphics[width=8.8cm]{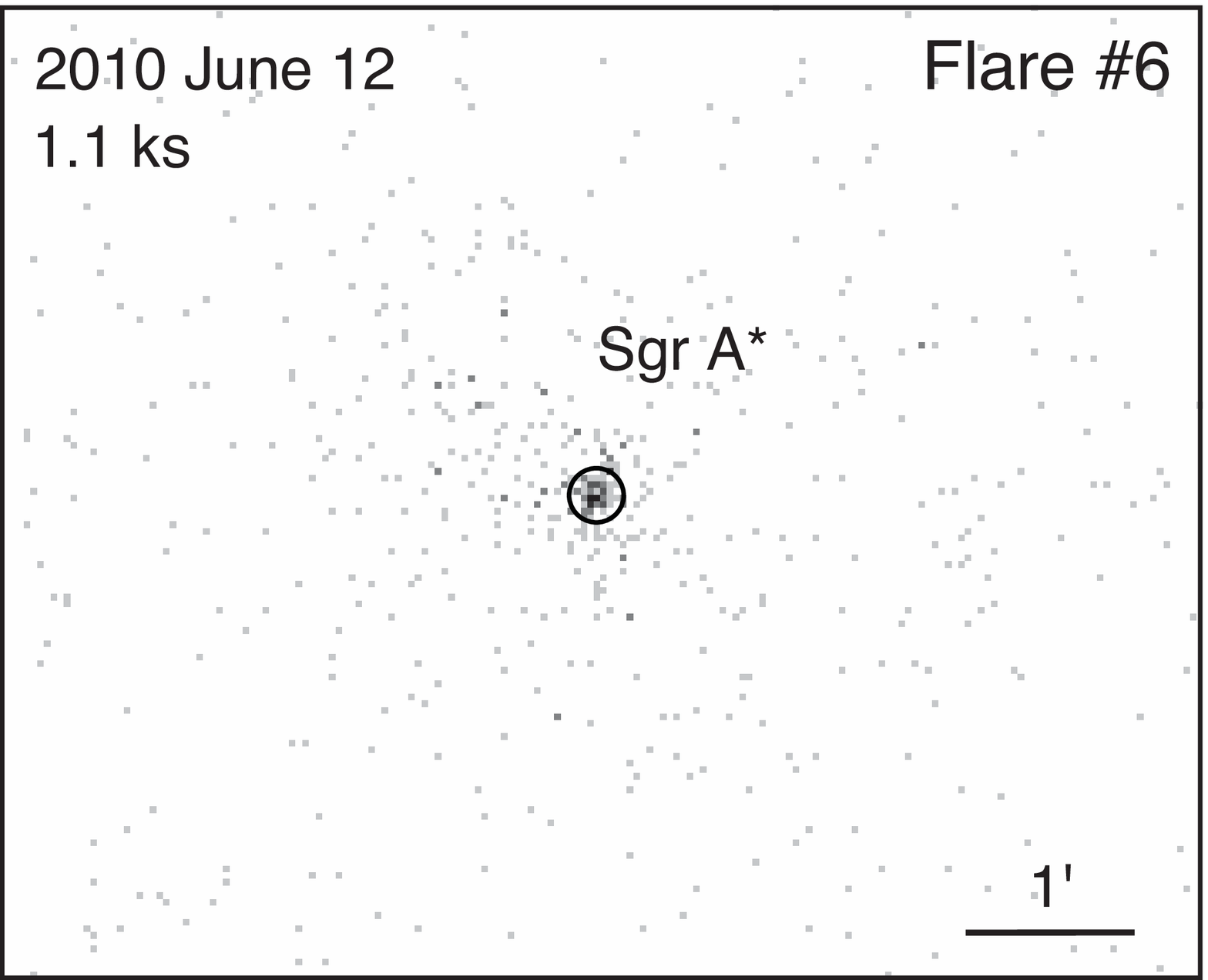}
    \end{center}
    \caption[]{XRT images (0.3--10 keV) illustrating the flare detected from \sgra\ on 2010 June 12 (\#6). The observation containing the flare (Obs ID 90416021) and the proceeding observation (Obs ID 90416020) had similar exposure times of $\simeq$1.1 and $\simeq$1.0~ks, respectively. The circle indicates the $10''$ extraction region that was used in our analysis of X-ray flares from \sgra.}
 \label{fig:imageflare}
\end{figure*}

The sixth X-ray flare was detected in an observation that had a total exposure time of $\simeq$1.1~ks and consisted of two orbits. The light curve of this observation is shown in Figure~\ref{fig:lcflare5}. Figure~\ref{fig:imageflare} compares the XRT image of the 1.1 ks flare observation with that of the proceeding 1.0 ks pointing, which was taken 3 days earlier. 

Enhanced activity is clearly detected in the first orbit (exposure time of $\simeq$0.5~ks), during which the intensity of \sgra\ is $\simeq1.0\times10^{-1}~\cnts$, i.e., a factor of $\simeq$10 ($\simeq 13.5 \sigma$) above the continuum count rate. In the second orbit ($\simeq$0.6~ks) that occurred $\simeq1.4$~h after the first, the count rate from \sgra\ was still $\simeq 2.5 \sigma$ above the mean intensity ($\simeq2.8\times10^{-2}~\cnts$). Comparing the XRT image of the second orbit with observations of similar exposure (i.e., $\simeq$0.6~ks), reveals an excess of photons that suggests that the flare was still detected in this gti. 

The intensity averaged over the full observation (i.e., 2 orbits) is $\simeq6.1\times10^{-2}~\cnts$ and a total of 65 counts were obtained. The average X-ray spectrum is plotted in Figure~\ref{fig:spec}, along with the spectral fit. 
The first and the second orbit collected 48 and 17 photons, respectively, which allows to fit the two spectra separately. The intensity inferred from the second orbit was a factor $\simeq3$ lower than during the first. This suggests that the second orbit likely caught the tail of the X-ray flare, whereas the first may have detected it closer to the peak. Spectral analysis suggests that both intervals had similar spectra, which is supported by the inferred hardness ratios (Table~\ref{tab:spec}). There is thus no indication of spectral evolution during the flare, although the errors bars are large.

%%%%%%%%%%%%%%%%%
% BRIGHT FLARE SUMMARY
%%%%%%%%%%%%%%%%%

\subsection{The Source of the Swift X-Ray Flares}\label{subsec:flaresum}
We determined the XRT position of the six confirmed X-ray flares utilizing the point spread fitting method of the algorithm developed by \citet{evans09}.\footnote[8]{Available at: http://www.swift.ac.uk/user$\_$objects/} The coordinates of the individual flares are all consistent within their 90\% confidence errors of $\simeq3.6''-3.9''$. The \swift\ positions are offset by $\simeq1.9''-5.6''$ from the radio position of \sgra\ \citep[][]{yusefzadeh1999}. Stacking the data of the six flares to improve the statistics yields R.A. = 17:45:40.13, decl. = --29:00:27.0 (J2000), with a 90\% confidence uncertainty of $3.5''$. This localization is within $\simeq 1.6''$ from the coordinates of the radio counterpart of \sgra\ and fully consistent within the errors.

Given the spatial resolution of the XRT, we cannot formally exclude that the six X-ray flares were caused by another X-ray source located within our $10''$ extraction region. However, long-term \chan\ monitoring has shown that all the faint persistent X-ray sources located within this region display a stable X-ray luminosity; none of them is known to exhibit X-ray variation of the observed magnitude \citep[][]{muno2009}. 

Objects such as accreting white dwarfs or X-ray binaries may display transient X-ray events, but these typically last much longer (days--months) than the short (hours) events that we detect. There is one known transient X-ray binary located within $10''$ from \sgra \citep[][]{porquet05_eclipser,muno05_apj633,muno05_apj622}. However, its position is $4.2''$ away from the XRT position, outside the 90\% confidence error, inferred for the six flares \citep[][]{muno05_apj633}. 

The above considerations suggest that it is unlikely that another X-ray source was responsible for causing the observed flares. The spectral parameters and intensities observed with the XRT are very similar to the X-ray flares seen with \xmm\ and \chan\ \citep[e.g.,][]{baganoff2001,goldwurm2003,belanger2005,porquet2003,porquet08,nowak2012}. This further strengthens the association with \sgra. We therefore conclude that we have detected six new bright X-ray flares from \sgra\ with \swift. 

\subsection{Different Types of X-Ray Flares?}\label{subsec:flarecomp}
\begin{figure}
 \begin{center}
	\includegraphics[width=8.8cm]{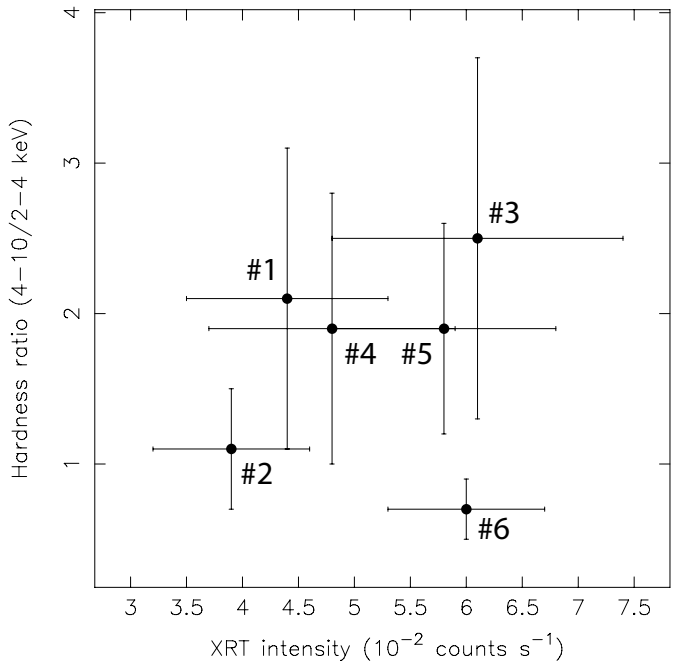}
    \end{center}
    \caption[]{\swift/XRT hardness ratio $HR$ (4--10 keV/2--4 keV) versus intensity for the six X-ray flares.
    }
 \label{fig:hr}
\end{figure}

The hardness ratios of the flares suggest that flare \#6 may have been spectrally softer than the other five (Figure~\ref{fig:hr}). To investigate whether the spectral differences could be biased by statistics, we co-added the spectra and weighted response files of flares \#1--5. Combining these yields 122 counts and a net count rate of $\simeq4.5\times10^{-2}~\cnts$. We obtain a photon index of $\Gamma=2.0\pm0.6$ and $HR=1.7\pm0.2$. This is indeed harder than flare \#6 ($\Gamma=3.0\pm0.8$ and $HR=0.7\pm0.2$), while their 2--10 keV absorbed fluxes are similar (Table~\ref{tab:flares}). 

\begin{figure}
 \begin{center}
	\includegraphics[width=8.8cm]{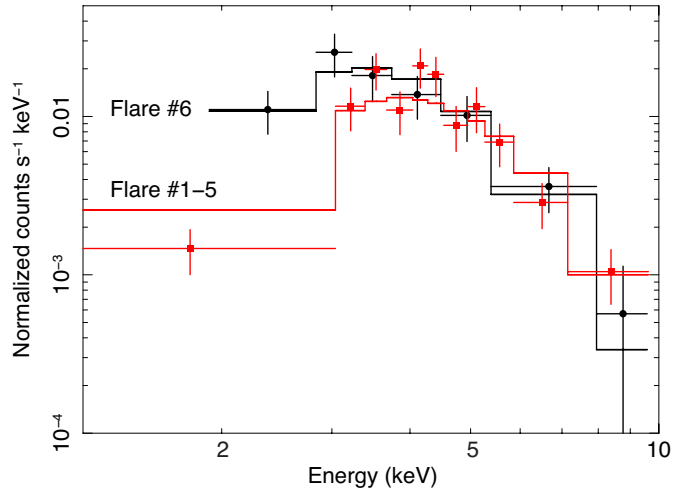} 
    \end{center}
    \caption[]{XRT spectrum of the flare detected from \sgra\ on 2010 June 12 (\#6; black bullets), compared to the summed spectra of flares \#1--5 (red squares). For representation purposes, the spectral data was rebinned to contain 10 photons per bin. The solid lines indicate fits to a combination of two absorbed power laws, with all parameters for the one representing the continuum emission fixed (see Section~\ref{subsec:specana})}.
 \label{fig:spec}
\end{figure}

Figure~\ref{fig:spec} compares the averaged spectrum of flares \#1--5 to that of flare \#6. This shows that the main difference between the spectra occurs at energies $\lesssim$3~keV. This suggests that possibly the absorption or dust scattering along our line of sight was different during flare \#6. Comparing the continuum spectra at different epochs does not indicate that the background emission was different at that time. We also fitted the spectrum of flare \#6 while subtracting the spectrum of the preceding observation (Obs ID 90416020) as a background. This yielded an index of $\Gamma = 2.6 \pm 1.0$ (with $N_{\mathrm{H}}=(9.1 \pm 0.5)\times10^{22}~\nh$ fixed; Section~\ref{subsec:continuum}). 
While this remains softer than the average value inferred for flares \#1-5, we cannot exclude that the apparent differences are caused by changes in the background emission.

Flare \#6 stands out in the long-term XRT light curve as the brightest event (Figure~\ref{fig:lc}), although it was detected in two orbits and its averaged intensity was similar to that of the other five flares (Table~\ref{tab:flares}). Spectral differences may possibly arise if we caught flare \#6 near its peak, whereas the other five events trace the rising or decaying parts of the flares. However, we did not find signs of spectral evolution during flare \#6 \citep[detailed analysis of other bright X-ray flares detected with \chan\ and \xmm\ did not reveal evidence for spectral evolution either;][]{porquet08,nowak2012}. Therefore, we do not consider it likely that the different spectral properties of the \swift\ flares are the result of tracing different parts of the X-ray flares.

The position of flare \#6 (R.A. = 17:45:40.27, decl. = --29:00:24.1 with a 90\% confidence error of $3.6''$) is $\simeq5''$ offset from the radio position of \sgra, but consistent with the location of the other five \swift\ flares. We do not consider it likely that flare \#6 was caused by a different source, although we cannot formally exclude this possibility (Section~\ref{subsec:flaresum}).

%%%%%%%%%%%%%%%
% DISCUSSION
%%%%%%%%%%%%%%%

\section{Discussion}\label{sec:discuss}
We have used six years of \swift/XRT monitoring data of the Galactic center to search for X-ray flares from the supermassive black hole. Within a $10''$ extraction region centered on \sgra, we detect a continuum emission level of $\simeq1.1\times10^{-2}~\cnts$. The inferred corresponding 2--10 keV luminosity is $L_{\mathrm{X}} \simeq 2 \times10^{34}~\lum$. This is a factor of $\simeq10$ above the quiescent X-ray emission of the supermassive black hole and is dominated by contributions from diffuse emission and faint X-ray point sources. 

We searched the long-term \swift\ light curve of \sgra\ for occurrences during which the XRT intensity increased a factor $>$3$\sigma$ above the mean (i.e., above a count rate of $\simeq3\times10^{-2}~\cnts$). This chosen threshold makes us sensitive to X-ray flares that reach a 2--10 keV intensity of $L_{\mathrm{X}} \gtrsim 7 \times10^{34}~\lum$ (i.e., a factor of $\gtrsim35$ above the quiescent level). This is at the high end of the luminosity distribution of those observed with \chan\ and \xmm, implying that we are sensitive only to the brightest X-ray flares from \sgra\ \citep[e.g.,][]{baganoff2001,baganoff2003,belanger2005,porquet2003,porquet08,nowak2012}.

Investigation of the images, light curves and spectra allowed us to identify six X-ray flares from \sgra. They have inferred 2--10 keV mean luminosities of $L_{\mathrm{X}}\simeq(1-3)\times10^{35}~\lum$ (Table~\ref{tab:flares}). Only four of such similarly bright events have been reported before: two were seen with \chan\ \citep[][]{baganoff2001,nowak2012} and two with \xmm\ \citep[][]{porquet2003,porquet08}. Using $\simeq0.8$~Ms of data, the \swift\ monitoring campaign has thus more than doubled the number of bright (i.e., $L_{\mathrm{X}} > 10^{35}~\lum$) X-ray flares detected from \sgra, setting the current counter to 10. 

The multitude of observations provided by the \swift\ campaign clearly has the advantage of increasing the detection likelihood of X-ray flares. This is also illustrated by the fact that we detected five X-ray flares between 2006 and 2008 when the Galactic center was observed every day, while only one flare was detected in 2009--2011 when the cadence was lowered to one pointing every three days (see Section~\ref{subsec:rate}).

\subsection{Properties of the X-Ray Flares Detected with Swift}\label{subsec:flareprop}
Prior studies with \chan\ and \xmm\ suggested that the brightest ($L_{\mathrm{X}} \gtrsim10^{35}~\lum$) X-ray flares had different spectral slopes \citep[cf.][]{baganoff2001,porquet2003,porquet08}. However, it was recently established that these differences were likely due to instrumental effects (primarily pile-up) and spectral modeling \citep[][]{nowak2012}. 

The \swift\ campaign has detected multiple bright X-ray flares with the same instrument. This allows for a comparative study of their properties unbiased by instrumental effects, which are induced when comparing flares detected with different instruments \citep[][]{porquet08,nowak2012}. The photon index inferred from \swift-flares \#1--5 ($\Gamma = 2.0\pm0.6$) is comparable to that of the bright flares detected by \chan\ and \xmm\ \citep[$\Gamma = 2.0^{+0.7}_{-0.6}$, $2.3\pm0.3$, and $2.4^{+0.4}_{-0.3}$;][]{nowak2012}. However, \swift-flare \#6 was softer ($\Gamma = 3.0\pm0.8$). 

Although we cannot exclude that this event was caused by a different source or that changes in the background emission play a role (Section~\ref{subsec:flarecomp}), the \swift\ results may indicate that flares of similar intensity can have different spectral shapes. This could possibly be explained in terms of different emission mechanisms (e.g., synchrotron versus inverse Compton radiation), different emitting regions (e.g., local versus global events), or different Lorentz factors. One model that can account for different spectral properties is the stochastic acceleration formalism proposed by \citet{liu2004}. Within this framework, harder flares can be produced by local magnetohydrodynamic processes such as magnetic reconnection, whereas softer flares are expected to result from global fluctuations.

\subsection{Constraints on the Flaring Repetition}\label{subsec:rate}
We can use the results of the \swift\ campaign to estimate the occurrence rate of bright X-ray flares from \sgra, i.e., those that have a luminosity of $L_{\mathrm{X}}\gtrsim 10^{35}~\lum$ (2--10 keV). Between 2006 February 24 and 2011 October 25, the Galactic center was observed with the XRT on 715 days. A total of six X-ray flares were detected during those observations. We cannot put firm constraints on the length the flares from the \swift\ data, but we will assume that the duration of all events was the same as similarly bright flares observed with \chan\ and \xmm, i.e., $\simeq$1--3~h \citep[][]{baganoff2001,porquet2003,porquet08,belanger2005,nowak2012}. 

Making the simplest assumption that the occurrence is uniform, we can estimate the flaring rate by dividing the number of events ($N$) by the number of days that was observed ($N_{\mathrm{obs}}$=715), and multiplying this with the time in a day ($t_{\mathrm{d}}$=24~h) divided by the duration of a flare ($t_{\mathrm{fl}}$=1--3~h). Thus, we calculate the number of flares per day as ($N$/$N_{\mathrm{obs}}) \times (t_{\mathrm{d}}$/$t_{\mathrm{fl}}$). For a total of six X-ray flares the flaring rate is then $\simeq$0.1--0.2~day$^{-1}$. This implies that a bright flare with an intensity of $L_{\mathrm{X}}\gtrsim 10^{35}~\lum$ occurs approximately every 5--10 days. We note that this flaring rate is consistent with the fact that several more flares were detected with \swift\ in 2006--2008 than in 2009--2011 (Figure~\ref{fig:lc}). This difference is simply explained by the larger number of observations (i.e., the higher sampling rate) during the first three years of the campaign (Table~\ref{tab:obs}).

The flaring rate that we obtain from the \swift\ data is in line with previous estimates. Based on 0.6~Ms of \chan\ data, \citet{baganoff2003_HEAD} infers an occurrence rate of $\simeq$0.3--0.9~flares~day$^{-1}$ for events that increase a factor $>$10 above the quiescent luminosity of \sgra. \chan\ has also exposed several faint X-ray flares that do not become brighter than $L_{\mathrm{X}}\lesssim 10^{34}~\lum$, and are thought to occur more frequently \citep[e.g.,][]{baganoff2003,eckart2004,eckart2006,eckart2008}. However, such weak flares are beyond the reach of \swift.

\subsection{Outlook}
It was recently discovered that an ionized gas cloud with an estimated mass of $\simeq$3~$\Mearth$ is rapidly moving towards \sgra, and is expected to make a close passage around 2013 September \citep[][]{gillessen2012,gillessen2012_2}. The cloud is expected to become disrupted due to its gravitational interaction with the supermassive black hole. Parts of the shredded gas may become accreted, enhancing the persistent emission of \sgra\ for the next few years--decades \citep[e.g.,][]{anninos2012,burkert2012,gillessen2012_2,moscibrodzka2012,schartmann2012}. 

Possibly, the interaction of the gas cloud with the shock at the Bondi-radius can generate instabilities that produce more powerful and more frequent X-ray flares of \sgra. Having mapped out the occurrence rate and X-ray properties of the flares detected with \swift\ in 2006--2011, provides a calibration point that allows to study whether the flaring rate of the supermassive black hole will be affected as a result of the approaching gas cloud.

%%%%%%%%%%%%%%%
% ACKNOWLEDGEMENTS
%%%%%%%%%%%%%%%

\acknowledgments
N.D. is supported by NASA through Hubble Postdoctoral Fellowship grant number HST-HF-51287.01-A from the Space Telescope Science Institute, which is operated by the Association of Universities for Research in Astronomy, Incorporated, under NASA contract NAS5-26555. R.W. is supported by a European Research Council starting grant. This work made use of public data from the \swift\ data archive, and data supplied by the UK \swift\ Science Data Center at the University of Leicester. \swift\ is supported at Penn State University by NASA contract NAS5-00136. This research has made use of the XRT Data Analysis Software (XRTDAS) developed under the responsibility of the ASI Science Data Center (ASDC), Italy. 

{\it Facility:} \facility{Swift (XRT)}

%%%%%%%%%%%%%%%
% REFERENCES
%%%%%%%%%%%%%%%

\end{document}